\documentclass{mem}
\usepackage{txfonts}\usepackage{balance}
\usepackage{graphicx}
\usepackage[a4paper,breaklinks,dvipdfm]{hyperref}

\begin{document}

\title{Microlensing, Brown Dwarfs and GAIA}

\subtitle{}

\author{N.~W.\ Evans}

\institute{
Institute of Astronomy,
Madingley Road,
Cambridge,
CB3 0HA
\email{nwe@ast.cam.ac.uk}
}

\authorrunning{Evans}

\titlerunning{GAIA Microlensing}

\abstract{The GAIA satellite can precisely measure the masses of
  nearby brown dwarfs and lower main sequence stars by the
  microlensing effect. The scientific yield is maximised if the
  microlensing event is also followed with ground-based telescopes to
  provide densely sampled photometry. There are two possible
  strategies.  First, ongoing events can be triggered by photometric
  or astrometric alerts by GAIA. Second, events can be predicted using
  known high proper motion stars as lenses. This is much easier, as
  the location and time of an event can be forecast. Using the GAIA
  source density, we estimate that the sample size of high proper
  motion ($>300$ mas yr$^{-1}$) brown dwarfs needed to provide
  predictable events during the 5 year mission lifetime is
  surprisingly small, only of the order of a hundred.  This is comparable
  to the number of high proper motion brown dwarfs already known from
  the work of the UKIDSS Large Area Survey and the all-sky WISE
  satellite. Provided the relative parallax of the lens and the
  angular Einstein radius can be recovered from astrometric data, then
  the mass of the lens can be found. Microlensing provides the only
  way of measuring the masses of individual objects irrespective of
  their luminosity.  So, microlensing with GAIA is the best way to
  carry out an inventory of masses in the solar neighbourhood in the
  brown dwarf regime.  }
\maketitle{}

\section{Introduction}

The GAIA satellite was successfully launched on 19th December 2013.
It reached the nothingness at the L$_2$ Lagrange point of the Earth
and Sun on 14th January 2014. After testing and commissioning, GAIA
will survey the sky from this vantage point. It will observe each
object on average 72 times during the five-year mission lifetime (Eyer
et al. 2013). GAIA will perform multi-epoch, multi-band photometry and
spectroscopy, but the uniqueness of GAIA lies in its astrometric
capabilities. The astrometric precision of GAIA depends on source
magnitude and allows stellar parallaxes to be measured with errors in
the range 10 to 300 $\mu$as for stars with magnitude in the range 12
to 20 (de Bruijne 2012).

Microlensing occurs when an object (lens) passes close to the line of
sight between observer and source. The light rays bend towards the lens
and distorted images are produced whose separation is too small to be
resolved by any telescope (see Evans 2003 or Mao 2012 for recent
reviews). Instead, there is a brightening of the source (photometric
microlensing) and displacement of the light centroid of the images
(astrometric microlensing). In principle, GAIA can detect both
photometric and astrometric microlensing, though the latter is a much
more attractive and feasible proposition.

Many thousand microlensing events have so far been identified,
primarily towards the Galactic Bulge. They have been found by
ground-based photometric surveys, the most active of which in recent
years are OGLE and MOA (see e.g., Sumi et al. 2013, Suzuki et al 2014,
Wyrzykowski et al. 2014). To date, no microlensing event has ever been
studied using both its astrometric and photometric signal, though this
can substantially break parameter degeneracies. GAIA will change all
this.

Microlensing is a unique technique, as it can measure the masses of
objects irrespective of their luminosity, irrespective of whether
they are bright or dark. It has become widely known as a tool for
exoplanet detection (e.g., Perryman 2011, Sumi et al. 2011), but it
has the capability to make substantial contributions to the study of
all faint stellar populations, such as brown dwarfs, white dwarfs and
neutron stars. It remains a technique whose time has yet to come. The
GAIA satellite with its powerful astrometric capabilities will usher
in the day.

\section{Preliminaries}

The all-sky averaged photometric microlensing optical depth is $\sim 5
\times 10^{-7}$. The typical duration of a photometric microlensing
event is 1-3 months, so there are a total of 8000-10000 photometric
events during the GAIA mission. GAIA's sampling law is peculiar, with
objects observed in clusters of 4-5 orbits, with gaps of 30-40 days
separating these groups of measurements. Most of the stars will have
on average about 70 measurements, though some areas of high star
density such as the Galactic Center will only have about 50
measurements.  GAIA photometric data alone will therefore rarely be
adequate to characterize events

Fortunately, there is also an astrometric microlensing signal
accompanying any photometric event. Although the two images of a
microlensed source are unresolvable, GAIA can measure the small
deviation (of the order of a fraction of a mas) of the centroid of the
two images. For microlensing, the increase in apparent brightness
falls off like (impact parameter)$^{-4}$, but the displacement in image
locations is proportional to (impact parameter)$^{-1}$. So, the
cross-section for astrometric microlensing is larger, as the effect is
a much more slowly declining function of impact parameter. Paczynski
(1996) already and clearly makes the point that astrometric
microlensing is a powerful way of measuring the masses of brown dwarfs
and other nearby stars, although he envisaged using the {\it Hubble
  Space Telescope} rather than GAIA. The astrometric cross-section is
proportional to the area a lens sweeps out on the sky, and so to the
product of lens proper motion and angular Einstein radius, which
favours nearby lenses.

Belokurov \& Evans (2002) showed that the all-sky averaged astrometric
optical depth is $\sim 2.5 \times 10^{-5}$. They carried out Monte
Carlo simulations assuming a distribution of lenses and sources from a
Galaxy model, together with the GAIA sampling algorithm. Since then,
there have been some modifications to the scanning law, though the
sampling properties do not change significantly. Allowing for the
astrometric precision, which is a function of the source magnitude,
they concluded that there are $\sim 15,000$ astrometric events in
which the centroid shift is greater than $5\sigma$ during the mission
lifetime. Some of these events cannot be identified, because any
identification algorithm will generate too many false positives.

For the study of dwarfs, the most important events are those for which
the mass of the lens is measurable. The mass is related to the angular
Einstein radius $\theta_{\rm E}$ and the relative parallax between
source and lens $\pi_{\rm sl}$ by (see eq. 40 of Dominik \& Sahu 2000)
$$
M = 0.125 M_\odot 
 \left({\theta_{\rm E} \over {\rm mas}} \right)^2 
\left( {\pi_{\rm sl} \over {\rm mas}} \right)^{-1}.
$$
Provided the astrometric fit enables $\pi_{\rm sl}$ and $\theta_{\rm
  E}$ to be extracted, then the lens mass is recoverable.  This
favours close lenses with larger Einstein radii and hence longer
timescales. Using a covariance analysis, Belokurov \& Evans (2002)
estimated that only for 10 \% of all astrometric events can the mass
of the lens can be recovered to good accuracy. This gives a sample of
$\sim 1,500$ `gold-plated' astrometric microlensing events in the GAIA
database for which all the microlensing parameters are obtainable.

So far, we have assumed that GAIA data alone may be enough to
characterize events. However, the number of `gold-plated' events can
be substantially increased if GAIA data are complemented with
ground-based photometry. The science alerts program (Wyrzykowski \&
Hodgkin 2012, Wyrzykowski et al. 2014b) has built an anomaly detection
pipeline which triggers on photometric brightening. GAIA observations
come in pairs separated by 106.5 minutes, which provides a check on
transients and a simple classifier.  Tests on SDSS Stripe 82 data
suggest that major types of transients -- such as microlensing -- can
be reliably identified and triggered with two datapoints. This raises
the prospect of follow-up photometry for on-going events, which will
substantially increase the numbers with accurate mass measurements.

\begin{figure*}[t!]
\resizebox{0.5\hsize}{!}{\includegraphics[clip=true]{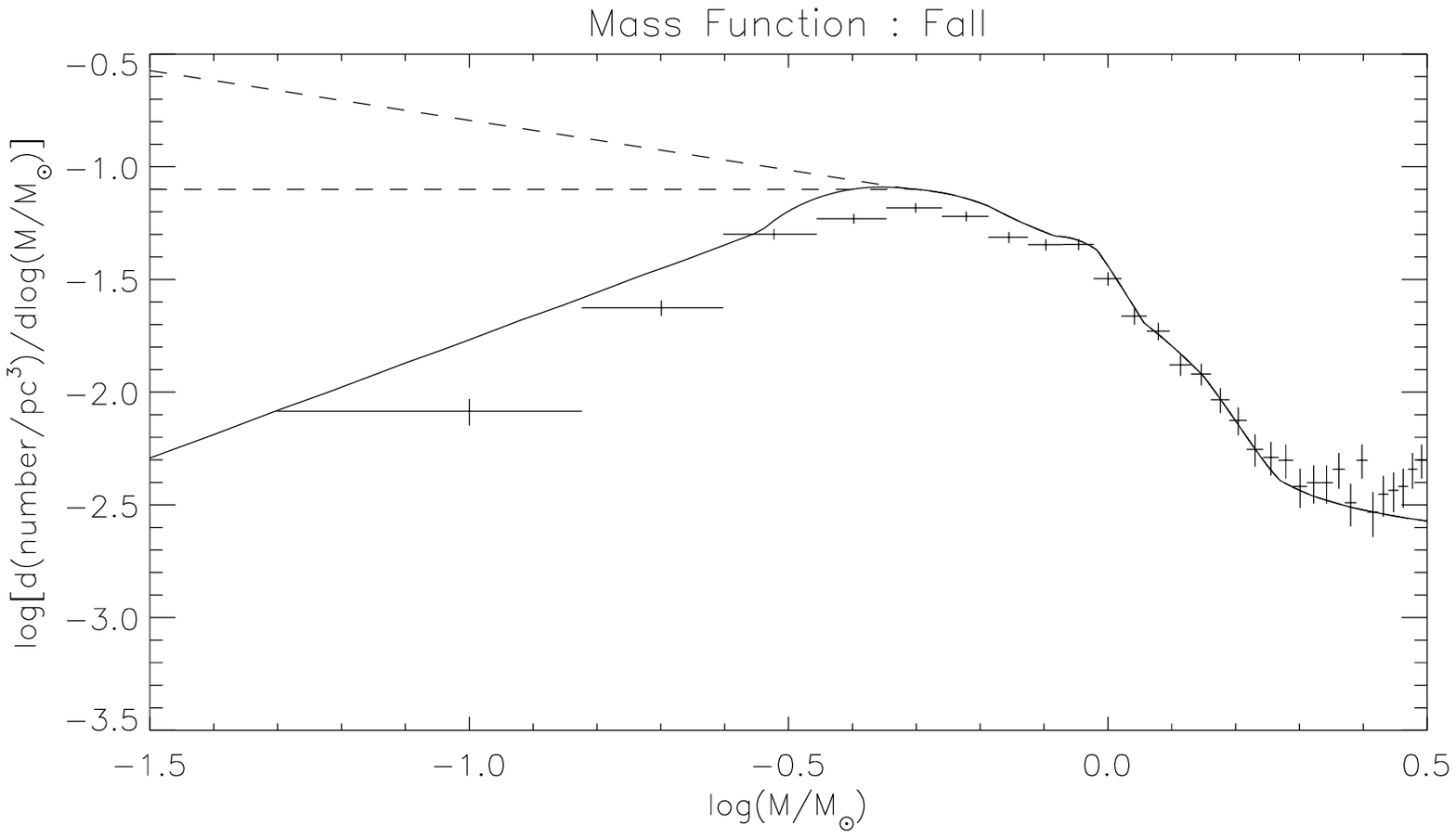}}
\resizebox{0.5\hsize}{!}{\includegraphics[clip=true]{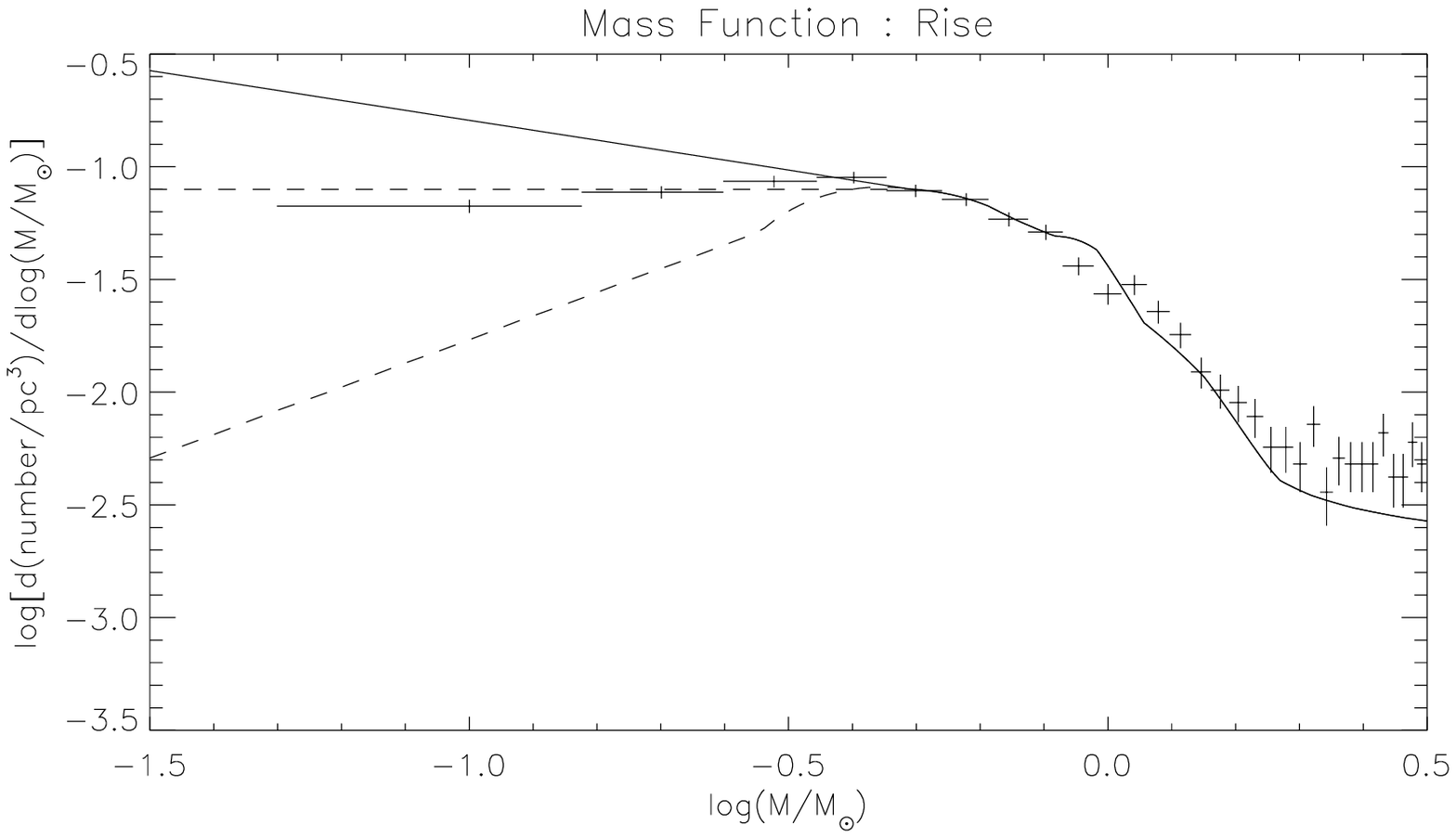}}
\caption{\footnotesize Comparison of the recovered mass function from
  GAIA's astrometric microlensing database with the true mass function
  in the case that it is falling (left) and rising (right). The full
  line shows the mass function used in Monte Carlo simulations to
  provide mock GAIA datasets. The datapoints show the estimates of the
  mass function from the $\sim 1500$ gold-plated events. This is discussed
  in more detail in Belokurov \& Evans (2002).}
\label{eta}
\end{figure*}

\section{The Local Mass Function}

The gold-plated events are wholly due to local lenses and this
suggests a natural application is to measure the local mass function.
Fig.~1, taken from Belokurov \& Evans (2002), shows the recovery of
the mass function in two cases in which it is assumed to be falling or
rising.  The mass functions are reproduced accurately above $0.3
M_\odot$. Below this, the mass functions fall below the true
curves. This is because astrometric microlensing events are biased
towards larger masses, which of course have larger Einstein radii.
Even so, the gold-plated events can distinguish between rising and
falling mass functions. Of course, the bias can be corrected for by
calibration against the simulations if we wish to reproduce the mass
function into the brown dwarf regime.

Local populations of low mass ($\sim 0.5 M_\odot$) black holes or
neutron stars could easily have eluded identification thus far.  Cool
halo white dwarfs are also extremely difficult to detect.  Existing
programs look for faint objects with high proper motions (e.g.,
Oppenheimer et al. 2001, Vidrih et al. 2007). Such searches miss stars
with low space motions and have difficulties detecting stars with
extremely high proper motions (depending on the epoch difference).
Experiments reveal that mass functions with spikes due to populations
of white dwarfs and neutron stars can be easily recovered.  In fact,
this is the regime in which GAIA's astrometric microlensing signal is
most efficient, as recently confirmed by Sajadian (2014). GAIA is the
first instrument to have the capabilities of detecting nearby
populations of very dark objects.

\section{Predictable Events}

The alternative to identifying events in GAIA's database is to
forecast events. A very attractive feature of astrometric microlensing
is that events can be predicted in advance, if the lens proper motion
is known (e.g., Paczynski 1995). For a brown dwarf at distance
$D_{\rm d}$ lensing a more distant source at $D_{\rm s}$, the angular
Einstein ring is
$$\left( \theta_{\rm E}\over {\rm mas}\right) = 8 \left( {M\over 0.08
  M_\odot} \right)^{1/2} \left(10 {\rm pc} \over D_{\rm d} \right)^{1/2} \left( 1 - {D_{\rm d}\over D_{\rm s}} \right)^{1/2}
$$
If the proper motion of the brown dwarf is ${\dot \theta}$, then
during the 5 year GAIA mission lifetime, it sweeps our an area on the
sky in square arcsec of
$$A \approx 0.024 \left( {M\over 0.08 M_\odot} \right)^{1/2} 
\left(10 {\rm pc} \over D_{\rm d}
\right)^{1/2}\left({{\dot \theta}\over 300 {\rm mas yr}^{-1}}\right)$$
The total number of stars detected by GAIA is not known, but it is
popularly supposed that GAIA will detect at least a billion stars.  If
so, the source density is $\sim 0.002$ stars per square arcsec.  For
comparison, the OGLE survey has a comparable limiting magnitude to
GAIA.  Towards the dense bulge star fields, the source density found
by OGLE is $\sim 0.15$ stars per square arcsec (Udalski et
al. 1994). However, even with the most pessimistic estimates, we still
expect a sample of a few tens of brown dwarfs with proper motions in
excess of 300 mas yr$^{-1}$ to yield events during the GAIA mission.
Not merely are the numbers very favourable, but there are other
advantages, too.  First, faint stars are favoured as lenses because
the light centroid shift of the source star is less affected than with
a bright lens. Second, close stars are favoured, as the centroid shift
is larger. So this is ideal for determining the masses of nearby brown
dwarfs and M dwarfs, provided we can identify enough to act as high
proper motion lenses in the first place.

Compared to alerts, this is a much easier strategy to follow as the
location and time of an event can be forecast ahead of time with fair
accuracy.  Of course, a predicted event can be followed with dense
ground-based photometry with which GAIA's astrometry can be combined
to yield very accurate mass measurements. Paczynski (1995) estimated
that in the most propitious cases an astonishing 1 \% error in the
mass is obtainable.  The difficulty lies in the fact that there is no
reliable all-sky catalogue of faint, high proper motion stars
available. Existing catalogues are partial and are often contaminated
by spurious entries arising from the digitalisation of old
photographic plates.  Nonetheless, some interesting results have been
obtained, albeit in a piecemeal manner.

Proft et al. (2011) searched though a number of proper motion
catalogues -- including LSPM-North of Lepine \& Shara (2005) and PPMX
of Roeser et al. (2008) -- for possible lenses that generate events
during the GAIA mission. Basically, this entails identifying
background stars that lie within a certain angular distance of the
future position of a high proper motion star. They identified nine
candidates which have a centroid shift between 100 and 4000 $\mu$as,
and argued that two of their candidates were exceptionally promising.
The first is the white dwarf LSPM JO431+5858E, which has a very high
proper motion of $2375$ mas yr$^{-1}$. It lenses a faint (19.7 mag)
background star with a time of closest approach of January 2014 ($\pm
1$ month). The deviation remains measurable for a further 100 ($\pm
20$) days. Unfortunately, given the delays in launch and in
commissioning caused by stray scattered light, GAIA will now only take
data well after the astrometric deviation has begun to subside.  The
second is the M dwarf LSPM J2004+3808 which has a proper motion of 341
mas yr$^{-1}$. It lenses a bright (12.7 mag) background star. The time
of closest approach is July 2014 ($\pm 6$ months) and the duration is
$2490 (\pm 1275)$ days. This event remains viable, and -- with a
predicted centroid shift of $\approx 1080$ $\mu$as -- should be easily
measurable by GAIA.

Another example was identified by Sahu et al. (2014). They found that
Proxima Centauri will pass close to a faint (19.5 mag) background star
in February 2016, giving a centroid shift of $\approx 1500$ $\mu$as.
Their motivation, though, is somewhat different, in that they suggest
that the astrometric signal may betray the presence of planets
(c.f.. Di Stefano et al. 2013).

So far, this technique has not been applied to measure brown dwarf
masses, yet it seems ripe for exploitation. Late-type M, L and T
dwarfs with high proper motion are beginning to be known in sufficient
numbers to make this feasible (e.g., Flaherty et al. 2009, Kirkpatrick
et al. 2010). The UKIDSS Large Area Survey has made -- and continues
to make -- striking contributions to the field, For example, Table 1
of Burningham et al. (2013) contains 28 T dwarfs with high proper
motion ($> 300$ mas yr$^{-1}$), whilst Table 1 of Smith et al.(2014)
gives a further 41 high proper motion late-type dwarfs. Overall,
scattered through the literature, there are probably already $\sim
1000$ known M, L and T dwarfs with high proper motion. Further, the
WISE satellite provides an all-sky and multi-epoch dataset, from which
color-selected brown dwarfs can be extracted with proper motions
(e.g., Kirkpatrick et al. 2011). This raises the prospect of a
full-sky volume-limited sample of high proper motion brown dwarfs,
which would be a gold-mine for prospecting for possible lenses for
GAIA. 

Finally, GAIA's early data releases will provide further candidate
high proper motion stars. For example, the Hundred Thousand Proper
Motion Catalogue (de Bruijne \& Eilers 2012) is scheduled for release
22 months after launch, and so in late 2015. These stars are in common
with Hipparcos and so have decade long baselines, though admittedly
only a handful are M dwarfs. The second GAIA catalogue is planned
for release 28 months after launch, and will provide the full
astrometric solution for all objects classified as single stars across
the whole sky. This of course will substantially increase the numbers
of high proper motion stars and can be used to predict events for the
latter half of the mission. Although the focus is on brown dwarfs
here, the masses of many classes of stars, such as bright supergiants
or cool white dwarfs, are not securely established and are interesting
targets in themselves.

\section{Conclusions}

The development of triggers for photometric microlensing has
significantly extended the microlensing capabilities of the GAIA
satellite (Wyrzykowski et al. 2014b). The first GAIA science alerts are
expected to be released in September 2014, before this article even
appears in print!  Belokurov \& Evans (2002) estimated that for $\sim
1500$ microlensing events, the mass of the lens could be deduced with
good accuracy from GAIA data alone.  However, follow-up photometry for
alerts, combined with GAIA's astrometric signal, substantially
increases the number of such `gold-plated' events, as well as the
accuracy of mass measurements. It may even be possible to alert for
microlensing events on astrometry, although the false positive rate
from binaries remains uncalculated and may be prohibitive.

The microlensing signal seen by GAIA is sensitive to local populations
of even the dimmest of stars and darkest of objects.  Astrometric
microlensing favours nearby lenses with high proper motion, as then
the deviation is greatest.  Objects with masses around $0.5$ to $1
M_\odot$, such as white dwarfs, black holes and neutron stars, are
particularly cleanly detected via astrometric microlensing. Brown
dwarfs have smaller Einstein radii and so the number of `gold-plated'
events is lower. The effect of this bias though is computable via
simulations and so one of the major scientific contributions that GAIA
can make is to determine the local mass function.

A much easier strategy than triggering events is predicting them.
Proft et al. (2011) have pioneered the prediction of astrometric
events for GAIA. Although one of the events that they predicted (white
dwarf LSPM JO431+5858E) is no longer viable given the delay in launch
and commissioning of GAIA, the other event (M dwarf LSPM J2004+3808)
remains feasible.  This though is a very attractive technique, and
there is scope for more work. We have estimated that a few tens of
high proper motion ($>300$ mas yr$^{-1}$) brown dwarfs are enough to
give astrometric lensing events during the GAIA mission. At first
glance, this is a surprisingly small number, but most of the optical
depth to astrometric microlensing is in nearby, fast-moving lenses.

The number of high proper motion brown dwarfs has increased by leaps
and bounds in recent years through surveys, such as UKIDSS. Building a
high proper motion all-sky catalogue from the multi-epoch WISE dataset
will yield many more brown dwarf lenses. Samples sizes of a hundred
should be enough to provide many predictable events during the GAIA
mission. Second, opportunities to forecast astrometric events in
advance are also provided by GAIA's early data releases. As currently
scheduled, 28 months after launch (that is, in April 2016), there will
be a data release of the full astrometric solution for single stars,
which can be mined for events in the latter half of the mission.

Of course, once the date of an event is known in advance, it can be
followed from the ground or even with the {\it Hubble Space
  Telescope}. Complementing GAIA's astrometric data with photometry
offers the prospect of very accurate mass determinations (to perhaps 1
\% in the most propitious cases), which will be invaluable for a range
of stars, not just the brown dwarfs considered in this article.

\begin{acknowledgements}
I thank Vasily Belokurov, Lukasz Wyrzykowski, Martin Smith and Ricky
Smart for many interesting conversations on this subject. It is a
pleasure to thank the Local Organizing Committee, and in particular
Ricky Smart, for a hugely enjoyable workshop.

\end{acknowledgements}

\bibliographystyle{aa}

\end{document}